\documentclass[12pt,preprint]{aastex}

\usepackage{emulateapj5}           
\usepackage{natbib}

\slugcomment{Draft version
  from \today. Accepted for publication in Astrophysical Journal Letters.}

\shorttitle{Kinematics of a hot massive accretion disk}
\shortauthors{Beuther \& Walsh}

\begin{document}

\title{Kinematics of a hot massive accretion disk candidate} 


\author{H.~Beuther$^1$ \& A.~Walsh$^{2}$}
\altaffiltext{1}{Max-Planck-Institute for Astronomy, K\"onigstuhl 17, 69117 Heidelberg, Germany}
\altaffiltext{2}{Centre for Astronomy, James Cook University, Townsville, QLD 4811 Australia}
\email{beuther@mpia.de,andrew.walsh@jcu.edu.au}

\begin{abstract}
  Characterizing rotation, infall and accretion disks around high-mass
  protostars is an important topic in massive star formation research.
  With the Australia Telescope Compact Array and the Very Large Array
  we studied a massive disk candidate at high angular resolution in
  ammonia (NH$_3$(4,4) \& (5,5)) tracing the warm disk but not the
  envelope.  The observations resolved at $\sim$0.4$''$ resolution
  (corresponding to $\sim$1400\,AU) a velocity gradient indicative of
  rotation perpendicular to the molecular outflow.  Assuming a
  Keplerian accretion disk, the estimated protostar-disk mass would be
  high, similar to the protostellar mass. Furthermore, the
  position-velocity diagram exhibits additional deviation from a
  Keplerian rotation profile which may be caused by infalling gas
  and/or a self-gravitating disk. Moreover, a large fraction of the
  rotating gas is at temperatures $>$100\,K, markedly different to
  typical low-mass accretion disks.  In addition, we resolve a central
  double-lobe cm continuum structure perpendicular to the rotation. We
  identify this with an ionized, optically thick jet.
\end{abstract}

\keywords{stars: formation -- stars: individual (IRDC\,18089-1732) --
  stars: early-type -- stars: rotation -- stars: winds, outflows}

\section{Introduction} 

Outflow and jet studies in high-mass star-forming regions have
revealed ample indirect evidence that accretion disks should be common
in these regions. However direct observational characterization of
such disks has so far been difficult (e.g.,
\citealt{beuther2006b,cesaroni2006}). While this is partly caused by
the typically large distances and the clustered mode of massive star
formation, an additional problem arises because these regions are
still deeply embedded within their natal cores, they have visual
extinctions of the order 1000, making it difficult to disentangle the
dense core emission from the genuine accretion disks. In the past,
different molecular line tracers were employed for different massive
star-forming regions (e.g., \citealt{cesaroni2006}), however, it is
interesting to note that for many sources only one or two molecular
lines exclusively indicated rotating motions whereas many other
promising spectral lines showed no such signatures (e.g.,
\citealt{beuther2007e}).  The problem of identifying a reliable
massive disk tracer prohibited statistical studies of larger disk
samples, until today.  Here we present new observations toward the
massive disk candidate IRAS\,18089-1732 in the high-excitation ammonia
lines NH$_3$(4,4) and (5,5), with lower energy levels of 200 and
295\,K, respectively.  While the high excitation temperatures ensure
that we are not tracing the cold gas envelope but only the warm
central regions (e.g., \citealt{cesaroni1998}), recent radiative
transfer calculations of 3D hydrodynamic massive core simulations
indicated that the wavelength regime $<100$\,GHz should be
particularly advantageous to studying the inner disk regions
\citep{krumholz2007a}. This work predicts high dust optical depths at
frequencies $>$100\,GHz potentially limiting spectral line
studies of the central disk in the (sub)mm regime. 

The target IRAS\,18089-1732 has previously been studied at (sub)mm
wavelengths with the Submillimeter Array, and a velocity gradient was
identified from east to west, perpendicular to the molecular outflow
emanating from the core approximately in north-south direction
\citep{beuther2004b,beuther2005c}. However, these data did not resolve
the velocity pattern of this structure and hence did not allow to
study its kinematic properties. The region is located at
$\sim$3.6\,kpc, it has a bolometric luminosity of about
$10^{4.5}$\,L$_{\odot}$, and it exhibits several other signs of early
massive star formation (e.g.,
\citealt{walsh1998,sridha,beuther2002a,beuther2002c,williams2004,williams2005,fuller2005,edris2007}).

\section{Observations}

\subsection{The Autralia Telescope Compact Array (ATCA)}

IRAS\,18089-1732 was observed in April 2007 with the ATCA in a 1.5\,km
baseline configuration including antenna 6. The phase reference center
was R.A. (J2000) $18^h11^m51^s.4$, Decl.~(J2000)
$-17^{\circ}31'28''.5$. We observed the NH$_3$(4,4) and (5,5)
inversion lines with the frequencies of the main hyperfine components
at 24.139 and 24.533\,GHz, respectively. The velocity relative to the
local standard of rest ($v_{\rm{lsr}}$) is $\sim 33.8$\,km\,s$^{-1}$.
Phase and amplitude were calibrated by regular observations of the
quasar 1829-207. Bandpass and flux were calibrated with observations
of 1741-038 and 1934-638. The spectral resolution of the observations
was 62\,kHz corresponding to a velocity resolution of $\sim
0.8$\,km\,s$^{-1}$.  Applying a robust weighting of 0.5, the $1\sigma$
rms per 0.8\,km\,s$^{-1}$ channel is $\sim$2.5\,mJy\,beam$^{-1}$. The
synthesized beam is $2.8''\times 0.4''$ with a position angle of
8.5$^{\circ}$. Because the outflow is in the north-south direction
\citep{beuther2004b} and the expected accretion disk in the east-west
direction, the small synthesized beam along the east-west axis is
fortunate for the analysis.

\subsection{The Very Large Array (VLA)}

To improve the beam shape, we observed the same region in October 2007
with the VLA/eVLA hybrid array in an extended configurations (BnA) in
the NH$_3$(5,5) line. Due to mediocre weather conditions and a worse
performance of the retrofitted eVLA antennas at high frequencies, we
could not usefully include the longest baselines, restricting the data
to uv-ranges $\leq$500\,k$\lambda$.  Bandpass and flux calibration
were conducted with 1331+305 and 1733-130. Fast switching between
source and phase calibrator 1832-105 were applied to calibrate the
phases and amplitudes. The original spectral resolution of 48.8\,kHz
was smoothed in the final data-cube to 1.0\,km\,s$^{-1}$.  Applying
natural weighting and restoring the data with a circular beam
(degraded to the larger axis of the fit to the dirty beam) the
synthesized beam of these data is $0.47''$. It was also possible to
extract a 1.2\,cm continuum map from the line-free part of the
spectrum. For this image we included longer baselines as well and the
synthesized beam is $0.26''\times0.19''$ with a PA of 84$^{\circ}$
from north.

\section{Results and discussion} 

\begin{figure}
\begin{center}
\includegraphics[angle=-90,width=17.5cm]{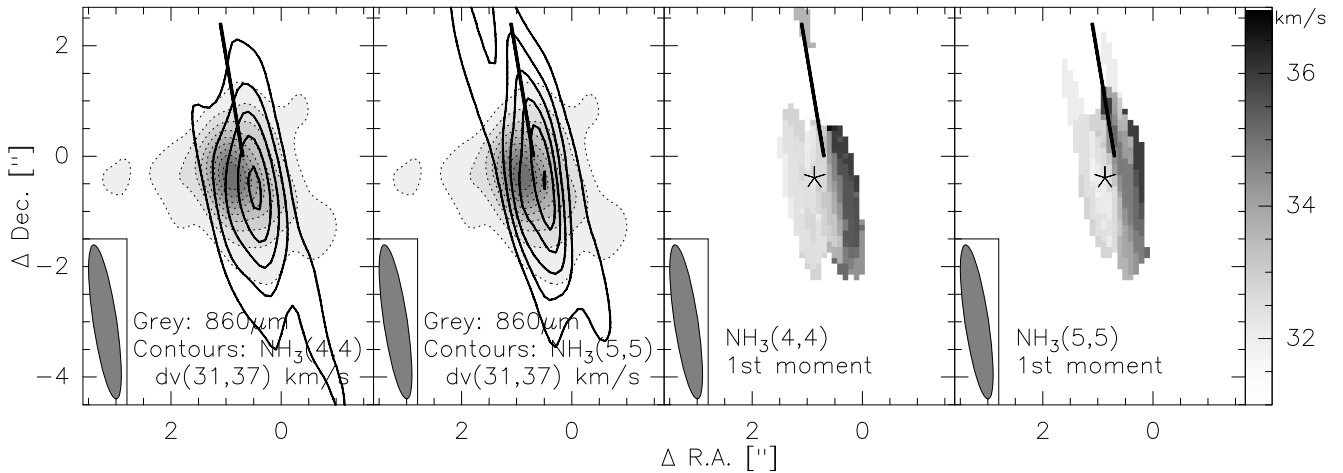}
\end{center}
\caption{The two left panels present in solid contours the NH$_3$(4,4)
  and (5,5) emission integrated from 31 to 37\,km\,s$^{-1}$.  The
  grey-scale shows the corresponding 860\,$\mu$m continuum emission
  \cite{beuther2005c}. The continuum emission is contoured from 10 to
  90\% (step 10\%) of the peak emission of 1.4\,Jy\,beam$^{-1}$. The
  NH$_3$ maps are contoured in $3\sigma$ steps with $3\sigma$ values
  of 3.75 and 3.3\,mJy\,beam$^{-1}$ for the (4,4) and (5,5) integrated
  maps, respectively. The two right panels show the corresponding 1st
  moment maps of the NH$_3$(4,4) and (5,5) spectral line data-cubes.
  The asterisks mark the peak position of the submm continuum
  emission, and the synthesized beam is shown at the bottom-left of
  each panel. The full line outlines the direction of the outflow as
  measured in SiO(8--7) \citep{beuther2005c}.}
\label{fig1}
\end{figure}

Figure \ref{fig1} presents a compilation of the ATCA data compared to
the submm continuum emission tracing the central dust and gas core
\citep{beuther2005c}. As expected the warm gas observed in the NH$_3$
lines traces the central core well. However, more interesting is the
velocity structure of the region. The two first moment maps (the
intensity weighted velocities) of the NH$_3$(4,4) and (5,5) data-cubes
both exhibit a velocity gradient from east to west, perpendicular to
the molecular outflow emanating approximately in north-south direction
\citep{beuther2004b}.  The north-south elongation seen in
Fig.~\ref{fig1} is caused by the non-symmetric beam shape. To improve
on that, we re-observed the source in the NH$_3$(5,5) line with the
VLA, Fig.\,\ref{fig2} presents and integrated and a first moment map
of these data.  The velocity gradient in east-west direction is
obvious again. The fact that the NH$_3$ structure is not flattened
could be either due to a viewing-angle of the rotating structure close
to face-on, or, since the NH$_3$ features in the north and south are
close to the $v_{\rm{lsr}}$, they may also contain contributions from
the ambient core.

In addition to the spectral line data, Fig.~\ref{fig2} presents the
simultaneously with the VLA observed 1.2\,cm continuum map. The cm
emission shows a double-lobe structure in north-south direction with
the dip centered on the submm continuum peak. The integrated emission
of this double-lobe is 7.4\,mJy. In combination with the previously
observed 3.6\,cm flux of 0.9\,mJy, the spectral index between these
two wavelengths is $\sim$2, consistent with an ionized, optically
thick jet \citep{reynolds1986}. Since the molecular outflow is also in
north-south direction we identify this double-lobe cm continuum
emission with the central jet perpendicular to the rotation axis of
the gas.

To get a better impression of the velocity field along the inferred
rotation axis, Fig.~\ref{fig3} shows a position-velocity digram cut
through the submm continuum peak position along the east-west axis.
For this analysis, we use the ATCA NH$_3$(4,4) data because they
combine the best signal-to-noise ratio with the highest spectral and
angular resolution along the given axis.  The velocity gradient is
clearly depicted here as well. With a spatial resolution of
$0.4''$along this axis, the structure with an approximate size of
$2''$ is well resolved and consistent with rotation of a potential
massive accretion disk. Furthermore, with a linear resolution of
$\sim$1400\,AU, these observations allow us, for the first time, to
better characterize the kinematics of the rotating structure.

Refering to the relatively simple picture of low-mass accretion disks
which are in Keplerian rotation (e.g., \citealt{ohashi1997,dutrey2007}),
we can calculate the rotationally supported binding mass
$M_{\rm{rot}}$ for IRAS\,18089-1732 assuming equilibrium between the
rotational and gravitational force at the outer radius of the disk.

\begin{eqnarray}
M_{\rm{rot}} & = & \frac{\delta v^2r}{G} \label{eq1} \\
\Rightarrow M_{\rm{rot}}[\rm{M_{\odot}}] & = &  1.13\,10^{-3} \times \delta v^2[\rm{km/s}]\times r[\rm{AU}] \label{eq2}
\end{eqnarray}

Here, $r$ is the disk radius ($\sim 1''\sim$3600\,AU), and $\delta v$
half the velocity regime observed in the first moment maps
($\sim$3\,km\,s$^{-1}$, Fig.~\ref{fig1}).  Equations \ref{eq1} \&
\ref{eq2} have to be divided by sin$^2(i)$ where $i$ is the unknown
inclination angle between the disk plane and the plane of the sky
($i=90^{\circ}$ for an edge-on system). With the given values, we can
estimate $M_{\rm{rot}}$ to $\sim 37$/(sin$^2(i)$)\,M$_{\odot}$. This
mass estimate is comparable to the core mass estimates from the submm
dust continuum emission (about 45\,M$_{\odot}$, \citealt{beuther2005c}).

A different way to assess potential (non-)Keplerian components in the
velocity structure is to draw an expected Keplerian velocity profile
for a 15\,M$_{\odot}$ protostar on the position-velocity diagram
(Fig.~\ref{fig3}, for the mass estimate see the following paragraph).
While for blue-shifted velocities and positive offsets the
position-velocity structure does approximately follow the Keplerian
profile, the red-shifted emission at negative velocities deviates more
strongly from the synthetic profile, even showing super-Keplerian
velocities in excess of the Keplerian speed. Furthermore, there are
additional components in the position-velocity digram, for example the
strong peak at (31\,km\,s$^{-1}$, $-0.2''$) that does not fit a
typical Keplerian disk. How can we explain the non-Keplerian
signatures in the rotating structure?

Even at the highest angular resolution (continuum and line data in
Fig.~\ref{fig1}), the source remains a single-peaked structure without
exhibiting signs of multiplicity. Although massive star formation
usually proceeds in a clustered mode, this indicates that most of the
bolometric luminosity of $10^{4.5}$\,L$_{\odot}$ likely stems from the
most massive central object. Assuming a main sequence star this would
give an upper mass limit of 20\,M$_{\odot}$ for the central object.
Since the source is likely still in its accretion phase, it has
additional accretion luminosity \citep{mckee2003,krumholz2006b} and
the actual protostellar mass may be lower. However, estimating for
example the accretion luminosity
$L_{\rm{acc}}=GM_*\dot{M}_{\rm{acc}}/R_*$ for a star of
$M_*=15$\,M$_{\odot}$ with a radius of $R_*=$10\,R$_{\odot}$ and an
accretion rate of $\dot{M}_{\rm{acc}}=10^{-4}$\,M$_{\odot}$yr$^{-1}$,
one derives an $L_{\rm{acc}}$ of only $\sim$5$\times
10^3$\,L$_{\odot}$. Although especially $\dot{M}_{\rm{acc}}$ and $R_*$
could both vary by a factor of a few (e.g., \citealt{krumholz2006b})
it is unlikely that the accretion luminosity contributes more than a
factor two to the overall luminosity. Therefore, we estimate the mass
of the central protostar to be likely between 15 and 20\,M$_{\odot}$.
Compared to the mass of the potential protostar/disk system calculated
either from the dust continuum emission or estimated via the
rotational support assumption, the actual protostellar mass is more
than a factor 2 lower. Hence, a Keplerian supported disk, where the
disk mass is negligible compared to the protostellar mass, is not
feasible in such a system. Because of the velocity gradient
perpendicular to the outflow (Figs.~\ref{fig1} \& \ref{fig2}) and the
comparatively low velocities compared to molecular outflows, it is
unlikely that these features are due to unbound motions, e.g., from
the molecular outflow. The more likely interpretation is that we are
dealing with a massive rotating structure not in Keplerian rotation.

Candidates for the non-Keplerian contributions are a rotating and
infalling envelope and a potentially self-gravitating disk structure.
For example, an infalling and rotating envelope produces similar
features in the position-velocity diagram like the blue-shifted peak
at (31\,km\,s$^{-1}$, $-0.2''$) in Fig.~\ref{fig3} (see, e.g., Fig.~10
in \citealt{ohashi1997}). Such rotating, infalling envelope structures
could resemble the proposed larger-scale toroids that are observed
toward a small number of even more luminous sources
\citep{cesaroni2005b,cesaroni2006}. It is expected that such toroidal
non-equilibrium structures should be gravitationally unstable, maybe
forming companion sources \citep{kratter2006,krumholz2006b}, which
then could produce additional kinematic signatures. In a similar
direction, \citet{keto2006} and \citet{keto2007} modeled the accretion
flow around high-mass stars: In their case, the gas orbits on
ballistic trajectories around a point mass (see also
\citealt{ulrich1976,terebey1984}), and the accretion flow appears
quasi-spherical at large distances from the star, flattening to a disk
at smaller radii due to conservation of angular momentum (see Fig.  9
in \citealt{keto2006}). Also noteworthy are the red-shifted
super-Keplerian gas velocities exceeding the Keplerian speed. While
sub-Keplerian disks are predicted for disks with significant
additional support against gravity than rotation (e.g., magnetic
field, \citealt{galli2006}), super-Keplerian velocities are expected
if the disk contributes a significant fraction of the mass to the
protostar-disk system. In this case, the outer disk rotates faster
because it feels the gravitational force of the protostar and the
inner disk (e.g., \citet{krumholz2006b} and Krumholz priv. comm.).

In addition to the kinematic analysis, ammonia is a well known
thermometer for molecular gas \citep{walmsley1983}. However, because
even these high-excitation lines still have high optical depths
($\tau>1$) we cannot accurately fit the hyperfine structure of the
lines.  Therefore, reasonable rotational temperature estimates are not
feasible. Nevertheless, the detection of spectral lines with
excitation temperatures $\geq 200$\,K implies that the average gas
temperatures should exceed 100\,K, consistent with previous
temperature estimates based on HCOOCH$_3$ and CH$_3$CN observations at
(sub)mm wavelengths \citep{beuther2004b,beuther2005c}. While the
previous (sub)mm temperature estimates were based on data that did not
resolve the kinematics of the rotating structure and therefore could
be attributed to the disk and/or envelope, the NH$_3$ data clearly
show that the central rotating structure harbors large amounts of warm
gas at temperatures $>$100\,K. This is markedly different to typical
low-mass disks where the bulk of the gas is at low temperatures
$<$30\,K \citep{pietu2007}.

\section{Conclusions and summary} 

From an observational point of view, this work shows that
high-excitation lines at 1.2\,cm wavelengths are well suited to study
the kinematics around massive protostars. While massive accretion disk
research in recent years has mostly focused on (sub)mm interferometer
observations, this may not be the ideal wavelength regime because of
high optical depth in the innermost regions \citep{krumholz2007a}.
Although there exist examples of NH$_3$ disk studies, these were
usually done in the low excitation lines NH$_3$(1,1) and (2,2) (e.g.,
\citealt{zhang1998a,beltran2006b}). Such low-energy line observations
suffer from contributions of the cold envelope material, and
high-excitation NH$_3$ observations like those presented here may turn
out to be one of the best tracers of the central accretion disks
around high-mass protostars.  Statistically larger samples have to be
investigated this way to base our knowledge on more solid ground.

In summary, these observations disentangle the kinematics, the mass
and the temperature of the rotating structure around a high-mass
protostar, identifying a hot massive accretion disk candidate.  The
kinematic signatures are different from typical Keplerian velocity
profiles known from low-mass star formation. These discrepancies could
be produced by larger-scale rotating and infalling toroids or
envelopes as well as by self-gravitating structures within the massive
accretion disk. The estimated gas temperature is $>$100\,K which is
also significantly larger than average temperatures from low-mass
accretion disks. While the angular resolution of $0.4''$ is very high,
corresponding to a projected linear resolution of $\sim$1400\,AU, we
are still not resolving the inner accretion disk regions.
Notwithstanding the fact that the larger-scale disk-like structure
observed here deviates from typical Keplerian disk structures, it may
be possible that such structures exist on even smaller scales closer
to the central protostar. Future observational tests include
high-spatial-resolution observations with ALMA and the completed eVLA,
as well as spatially and spectrally resolved observations of
vibrationally excited CO lines observable at mid-infrared wavelengths
(e.g., CRIRES on VLT).

\acknowledgments{We like to thank Fabian Walter and Hendrik Linz for
  their help with the VLA data. H.B. acknowledges financial support by
  the Emmy-Noether-Programm of the Deutsche Forschungsgemeinschaft
  (DFG, grant BE2578).  }


\begin{figure}
\begin{center}
\end{center}
\caption{The left panel shows the VLA NH$_3$(5,5) emission integrated
  from 31 to 37\,km\,s$^{-1}$ and contoured from 10 to 90\% (step
  20\%) of the peak emission. The right panel presents the
  corresponding 1st moment map contoured from 31.5 to
  36.5\,km\,s$^{-1}$ (step 1\,km\,s$^{-1}$).  The white-black dashed
  contours show the 1.2\,cm continuum emission. Contour levels start
  at $4\sigma$ and continue in $1\sigma$ steps of
  0.28\,mJy\,beam$^{-1}$.  The asterisks mark the position of the
  submm continuum peak \citep{beuther2005c}, and the synthesized beams
  are shown at the bottom-left (grey NH$_3$ and dashed 1.2\,cm
  emission).}
\label{fig2}
\end{figure}

\begin{figure}
\begin{center}
\includegraphics[angle=-90,width=8.4cm]{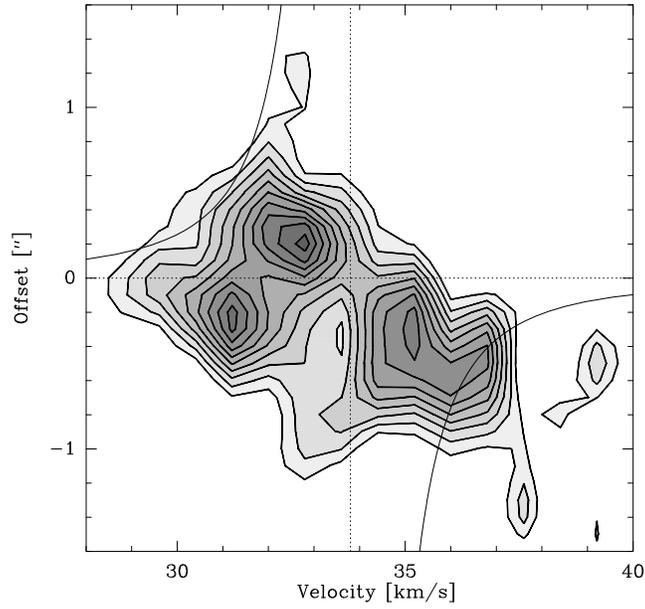}
\end{center}
\caption{Position-velocity diagram of the NH$_3$(4,4) data through the
  submm peak position in east-west direction. The contouring starts at
  the $2\sigma$ level and continues in $1\sigma'$ steps ($1\sigma\sim
  2.5$\,mJy\,beam$^{-1}$). The dotted lines mark the central position
  and the systemic velocity. The full line shows a Keplerian rotation
  curve with a central mass of 15\,M$_{\odot}$, which obviously fail
  to fit the data well.}
\label{fig3}
\end{figure}

\end{document}